\documentclass[%
 reprint,
 superscriptaddress,
 groupedaddress,
 amsmath,amssymb,
 aps,
 prb,
 floatfix,
]{revtex4-2}

\usepackage{graphicx}% Include figure files
\usepackage{dcolumn}% Align table columns on decimal point
\usepackage{bm}% bold math
\usepackage{lipsum} % generates random paragraphs
\usepackage{multirow}
\usepackage{threeparttable}
\usepackage{amsmath}
\usepackage{amssymb}
\usepackage{booktabs}
\usepackage{color}
\definecolor{myblue}{rgb}{0,0,1}

\usepackage[breaklinks=true,colorlinks=true,linkcolor=myblue,urlcolor=myblue,citecolor=myblue]{hyperref}% add hypertext capabilities

\usepackage{cleveref}

\begin{document}

\title{Predicting Methane Adsorption in Metal-Substituted MOFs: A Comparative Study between Density Functional Theory and Machine Learning}

\author{Karim Aljamal}
 \affiliation{Department of Chemistry and Biochemistry, University of California Santa Cruz, Santa Cruz, California, 95064 USA}
\author{Xiao Wang}%
 \email{xwang431@ucsc.edu}
 \affiliation{Department of Chemistry and Biochemistry, University of California Santa Cruz, Santa Cruz, California, 95064 USA}

\begin{abstract}
Metal-organic frameworks (MOFs) are promising materials for methane capture due to their high surface area and tunable properties. Metal substitution represents a powerful strategy to enhance MOF performance, yet systematic exploration of the vast chemical space remains challenging. In this work, we compare density functional theory (DFT) and machine learning (ML) in predicting methane adsorption properties in metal-substituted variants of three high-performing MOFs: M-HKUST-1, M-ATC, and M-ZIF-8 (M = Cu, Zn). DFT calculations reveal significant differences in methane binding energies between Cu and Zn variants of all three MOFs. On the other hand, we fine-tuned a pretrained multimodal ML model, PMTransformer, on a curated subset of hypothetical MOF (hMOF) structures to predict macroscopic adsorption properties. While the fine-tuned heat of adsorption model and uptake model qualitatively predict adsorption properties for original unaltered MOFs, they fail to distinguish between metal variants despite their different binding energetics identified by DFT. We trace this limitation to the hMOF training data generated using Grand Canonical Monte Carlo (GCMC) simulations based on classical force fields (UFF/TraPPE). Our study highlights a key challenge in ML-based MOF screening: ML models inherit the limitations of their training data, particularly when electronic effects at open metal sites significantly impact adsorption behaviors. Our findings emphasize the need for improved force fields or hybrid GCMC/DFT datasets to incorporate both geometric and electronic factors for accurate prediction of adsorption properties in metal-substituted MOFs.

\end{abstract}

\maketitle

%\tableofcontents

\section{Introduction}
\label{sec:intro}

As a greenhouse gas, methane (CH$_4$) is 28 times more effective than carbon dioxide (CO$_2$) at trapping heat in the atmosphere over a 100-year period.~\cite{ipcc2014} Despite its low atmospheric concentration (1929 ppb compared to 426 ppm of CO$_2$, measured in 2024~\cite{nasa2008}), CH$_4$ is responsible for 20-30\% of climate warming since pre-industrial times.~\cite{ipcc2023} Moreover, rapid expansion of unconventional oil and gas extraction has led to increased methane leakage, while the warming Arctic presents a looming threat of large-scale methane release.~\cite{stolaroff2012} Given methane's environmental impact, as well as its use as a cleaner energy source compared to oil and coal,~\cite{sapag2010} methods that can efficiently capture and store methane are of particular interest. Among current technologies, solid sorption using porous materials such as zeolites, activated carbons, and metal-organic frameworks (MOFs) are attractive due to broad applicability, energy efficiency, and environmental friendliness.~\cite{makal2012,tsivadze2018} Of these materials, MOFs have garnered particular attention in recent years. MOFs comprise metal ions or metal-containing clusters connected by organic ligands to form 3D networks, offering high surface area, adjustable pore size, and extensive chemical tunability. These features make MOFs promising candidates for methane adsorption.~\cite{lin2017,li2021,peng2013,he2014} Some notable names of high performing MOFs for methane adsorption include HKUST-1~\cite{wu2015a,denning2020,chong2023}, MOF-74~\cite{dietzel2009,wu2009,bao2011}, PCN-14~\cite{ma2008}, NU-111~\cite{peng2013a}, ZIF-8~\cite{casco2016,denning2021}, and MOF-5~\cite{saha2010}.

The inherent modularity of MOFs allows for systematic variation of their building blocks{\textemdash}similar to molecular Lego pieces{\textemdash}to tune their adsorption properties. A common strategy is substituting the metal centers in MOFs with alternative metals in order to alter their binding affinity, selectivity, and storage capacity for adsorbates. This approach is particularly effective for MOFs with coordinatively unsaturated metal centers (open metal sites) that can interact strongly with adsorbate molecules. A number of metal-substituted MOFs have been synthesized and characterized experimentally, showing enhanced performance.~\cite{uzun2014,han2014,queen2014,kapelewski2014,wasik2024,savagallis2015,wu2009,botas2010} For instance, 
Sava Gallis et al. showed that metal substitution of Cu in HKUST-1 with Mn, Fe, and Co increases the O$_2$/N$_2$ selectivity at 77K.~\cite{savagallis2015}
Similarly, Queen et al. demonstrated that substituting the open-metal site in M-MOF-74 [also known as CPO-27-M/M$_2$(dhtp)/M$_2$(dobdc); M refers to the choice of metal] leads to varying affinity of CO$_2$ as Mg $>$ Ni $>$ Co $>$ Fe $>$ Mn $>$ Zn $>$ Cu.~\cite{queen2014} 
In another popular system, MOF-5, Botas et al. found that partially substituting Zn by Co ions with no more than 25\% metal content systematically increases H$_2$, CH$_4$, and CO$_2$ storage capacity.~\cite{botas2010}
Metal substitution in MOFs creates a vast design space for tailoring MOFs to specific adsorption applications. 
However, with dozens of potential metals that could be incorporated into thousands of MOF structures, experimental screening becomes prohibitively resource-intensive. This is an ideal scenario for computational screening to identify promising candidates for methane capture and guide experimental efforts.

Computational methods have become invaluable tools for exploring the chemical space of metal-substituted MOFs and predicting their adsorption properties. 
Density functional theory (DFT) serves as the primary first-principles approach for accurately modeling the electronic structures, binding energetics, and adsorption mechanisms that govern MOF-adsorbate interactions.
Using PBE with dispersion correction (Grimme's DFT-D2~\cite{grimme2006}), Park et al. showed that Ti- and V-substituted MOF-74 exhibit enhanced CO$_2$ binding affinity compared to other M-MOF-74 variants (M = Mg, Ca, and the first transition metal elements).~\cite{park2012}
Lee et al. used vdW-DF2,~\cite{lee2010} a van der Waals density functional, to study the binding enthalpies of 14 small molecules (including methane) in M-MOF-74 (M = Mg, Ti, V, Cr, Mn, Fe, Co, Ni, Cu, Zn). Their calculated binding strengths that vary with the type of metal agree well with measured adsorption isotherms.~\cite{lee2015b}
Despite its accuracy, DFT's computational intensity typically limits its application to low adsorbate loading regimes ($\sim$1-10 guest molecules) or small-scale screening ($\sim$10-100 MOFs).
To predict macroscopic adsorption properties like volumetric uptake, heat of adsorption, and adsorbate diffusivity, DFT is commonly complemented by Monte Carlo (MC) or molecular dynamics (MD) approaches, which rely on force fields that must be carefully parameterized to reproduce experimental or ab initio quantum chemistry data.
For instance, Koh and coworkers leveraged vdW-DF2 interaction energies between CH$_4$ and open metal sites to fit force fields for 18 metal-substituted HKUST-1 variants. The new force fields were then used in Grand Canonical Monte Carlo (GCMC) simulations to predict methane isotherms at 298 K and yield excellent agreement with experiment.~\cite{kohPredictingMethaneStorage2015}
MC/MD simulations are often limited by the force field accuracy and the computational cost of parametrization.

In recent years, machine learning (ML) has emerged as a promising tool that can leverage existing computational/experimental data to rapidly predict methane adsorption properties across thousands of MOFs, showing the potential to drastically accelerate MOF screening and design processes.
Among the ML models developed specifically for MOFs, Liang et al. demonstrated that an XGBoost~\cite{chen2016a} model with only structural descriptors (such as surface area, pore volume, and density) could achieve high accuracy in predicting Xe/Kr adsorption properties.~\cite{liang2021a} Their model, trained on the Material Genomic MOFs (GMOF) database~\cite{lan2019}, achieved R$^2$ values exceeding 0.95 when benchmarked against GCMC simulations. More recently, the PMTransformer model introduced by Park et al.~\cite{park2023a} uses a multimodal Transformer architecture pretrained on 1.9 million hypothetical porous materials (including 1 million MOFs) to predict multiple properties simultaneously. This model captures complex chemical and spatial features of porous materials by processing them as atom-based graphs and 3D energy grids that map the interaction energy between material and a probe gas molecule. The fine-tuned PMTransformer achieved lower mean absolute errors than other state-of-the-art models (such as CGCNN~\cite{xie2018a} and MEGNET~\cite{chen2019}) across diverse porous material families and gas/property types.
Beyond these examples, other notable MOF-specific ML frameworks include MOF-NET~\cite{lee2021a}, MOFormer~\cite{cao2023}, Uni-MOF~\cite{wang2024}, and MSAIGNN~\cite{li2025}. Several universal models developed for crystalline materials, such as SchNet~\cite{schutt2018a}, MatFormer~\cite{yan2022}, and the aforementioned CGCNN~\cite{xie2018a} and MEGNET/M3GNET~\cite{chen2019,chen2022}, can also be applied for MOFs.
While these ML models have shown impressive performance for conventional MOFs, they have not been systematically tested on metal-substituted MOFs, which present a unique challenge due to non-trivial electronic effects of varied metal centers on adsorption properties. A simple question arises: can ML models trained on existing MOF databases correctly capture the trend in adsorption properties when metal centers are substituted in MOFs? This motivates the present study, where we benchmark the performance of ML and DFT methods in predicting methane adsorption properties of three high-performing MOFs (HKUST-1, Cu-ATC, and ZIF-8) and their metal-substituted variants.

The rest of the paper is organized as follows. In Section \ref{sec:comput_details}, we provide details on the MOF structures, computational methods, and simulation parameters used in this study. In Section \ref{sec:results}, we discuss DFT and ML results, followed by GCMC simulations that explain the limitations of ML. We conclude in Section \ref{sec:conclusions} with a summary of the findings and future directions.

\section{Computational Details}
\label{sec:comput_details}

\subsection{Metal-Organic Frameworks}
\label{ssec:comput_details_mofs}

\begin{figure*}[t]
    \includegraphics[width=0.9\textwidth]{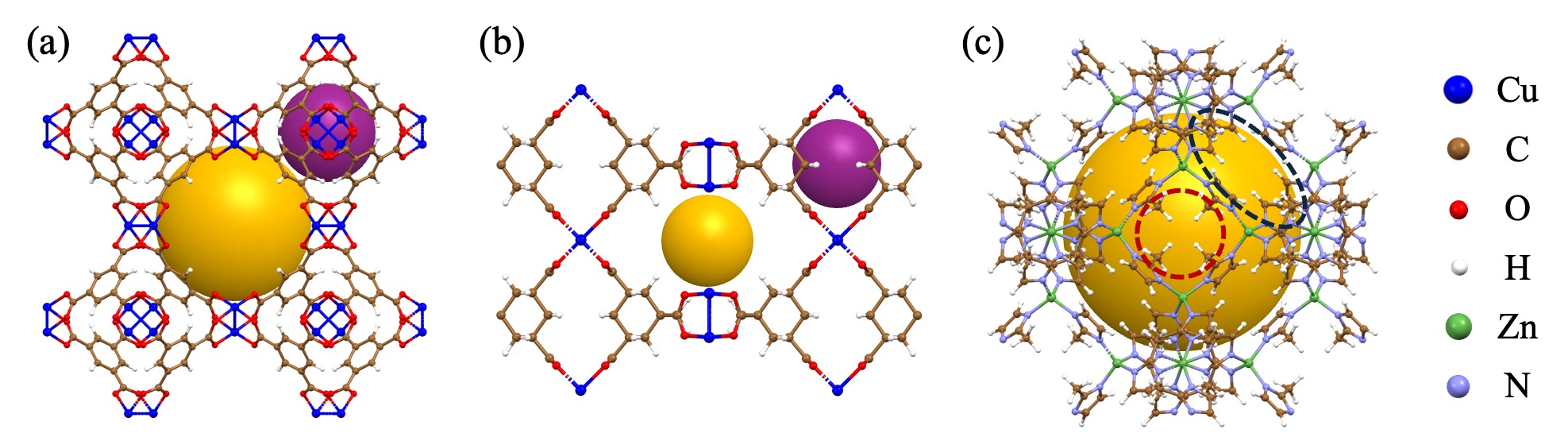}
    \caption{Bare MOF structures: (a) HKUST-1, (b) Cu-ATC, and (c) ZIF-8. The yellow and purple spheres represent major structural pores. The nano-trap in Cu-ATC is highlighted by the yellow sphere in (b). The 6-ring and 4-ring windows in ZIF-8 are marked with blue and red dashed circles, respectively.}
    \label{fig:bare_mofs}
\end{figure*}

In this study, we selected three typical MOFs with distinct structural characteristics to investigate the effects of metal substitution on methane adsorption: HKUST-1, Cu-ATC, and ZIF-8. HKUST-1 (also known as Cu-BTC; BTC = 1,3,5-benzene tricarboxylate) represents one of the most extensively studied MOFs for methane adsorption.~\cite{wu2015a,denning2020,chong2023} It is characterized by Cu paddlewheel units connected by BTC ligands, resulting a structure with large pores of 4-11 \AA\, (Fig. \ref{fig:bare_mofs}a). 
Cu-ATC (ATC = 1,3,5,7-adamantane tetracarboxylate) also features Cu paddlewheel units, but connected by ATC ligands (Fig. \ref{fig:bare_mofs}b). In 2019, Niu et al. found that the oppositely adjacent Cu paddlewheels in Cu-ATC create a methane nano-trap with the size of 4.43 \AA\, (yellow sphere in Fig. \ref{fig:bare_mofs}b), leading to a record-high methane uptake at 298 K and 1 bar.~\cite{niu2019}
In both HKUST-1 and Cu-ATC, the Cu$^{2+}$ center serve as an open-metal site, which is one of the primary binding sites for CH$_4$ due to enhanced Coulomb attraction. Another strong binding site is located at the organic linkers that form small pockets, where CH$_4$ can interact with the framework through van der Waals forces.
ZIF-8 (zeolitic imidazolate frameworks) is a sodalite-type structure with Zn centers tetrahedrally coordinated to imidazolate linkers and exhibits high methane adsorption capacity.~\cite{awadallah-f2019,denning2021} Unlike the other two MOFs, ZIF-8 lacks open metal sites but contains large pores of diameter 11.6 \AA\, connected through pore windows of 6-membered and 4-membered zinc rings (referred to as 4-ring windows and 6-ring windows; Fig. \ref{fig:bare_mofs}c),~\cite{bergaoui2021} which provide major binding sites for CH$_4$. This MOF is included to examine how metal substitution affects adsorption in structures without coordinatively unsaturated metal centers.

The initial structures of the three MOFs were obtained from experimentally determined crystallographic data reported in the literature.~\cite{getzschmann2010,niu2019,shekhah2014}
These structures were subsequently optimized using periodic DFT calculations (see details in Section \ref{ssec:comput_details_dft}). For each MOF, we created metal-substituted variants by replacing the original metal centers with an alternative metal. Specifically, we substituted Cu in HKUST-1 and Cu-ATC with Zn, and Zn in ZIF-8 with Cu. 
These substitutions were chosen because Cu and Zn have similar coordination environments in MOFs and can maintain the structural integrity of the system.
This creates a good stress test for ML models' ability to capture complex electronic effects rather than merely geometric factors.  
The metal-substituted structures were then reoptimized to relax atomic positions and lattice parameters. For clarity, we refer to the original MOFs as Cu-HKUST-1, Cu-ATC, and Zn-ZIF-8, and the metal-substituted variants as Zn-HKUST-1, Zn-ATC, and Cu-ZIF-8.

\subsection{Density Functional Theory}
\label{ssec:comput_details_dft}

Periodic density functional theory (DFT) calculations were conducted with the Quantum Espresso software package.~\cite{giannozzi2009,giannozzi2017} The Perdew-Burke-Ernzerhof (PBE)~\cite{perdew1996a,perdew1997} functional with Grimme's DFT-D3~\cite{grimme2010} dispersion correction was utilized for all energy and structural relaxation calculations. A plane wave basis set with projector augmented wave (PAW) pseudopotentials was used for all calculations. The kinetic energy cutoff was set to 1150 eV and the k-point mesh to N$\times$N$\times$N (N = 1, 5, and 3 for M-HKUST-1, M-ATC, and M-ZIF-8, respectively; M = Cu, Zn). The atomic positions and lattice parameters of both original and metal-substituted MOF structures were optimized using the Broyden-Fletcher-Goldfarb-Shanno (BFGS) algorithm. The convergence criteria for the self-consistent field (SCF) calculations were set to 1$\times$10$^{-8}$ Rydberg (1 Rydberg = 13.6 eV). The total energy and forces were converged to 1$\times$10$^{-4}$ Rydberg and 1$\times$10$^{-3}$ Rydberg/Bohr, respectively.

Structural relaxation of MOF-methane complex system was conducted for each MOF at two primary binding sites (see Section \ref{ssec:results_dft} for binding site details). These sites have been identified in previous studies~\cite{wuMetalOrganicFrameworks2010a,niu2019,fairen-jimenezFlexibilitySwingEffect2012} and are known to exhibit strong methane adsorption. Initial guesses of the binding structures were obtained using MOF Big Adsorbate Initializer (MBAI),~\cite{mbai-github} which guides the adsorbate molecule to the specified binding sites with Monte Carlo simulations through the RASPA software package.~\cite{dubbeldam2016} The binding sites were then optimized using PBE-D3 with the same settings as above to obtain the total energies of the MOF-adsorbate complexes, $E_{\text{complex}}$. The adsorption energy was calculated as:
\begin{equation}
\Delta E = E_{\text{complex}} - E_{\text{MOF}} - E_{\text{methane}},
\label{eq:adsorption_energy}
\end{equation}
where $E_{\text{MOF}}$, and $E_{\text{methane}}$ are the total energies of the bare MOF and the methane molecule, respectively, after geometry optimization. The adsorption energy provides a measure of the interaction strength between the MOF and methane, with a more negative value indicating a stronger binding affinity.

\subsection{Machine Learning Method}
\label{ssec:comput_details_ml}

We employed the PMTransformer model~\cite{park2023a} for predicting methane adsorption properties in MOFs. This model's architecture captures local atomic information and global framework geometry, both of which are essential for accurately modeling adsorption properties in MOFs, making it well-suited for our study. Furthermore, the model was pretrained on a diverse set of hypothetical porous materials, including 1 million MOFs, 
519,606 COFs (covalent organic frameworks), 277,250 PPNs (porous polymer networks), and 100,000 zeolites. Pre-training tasks include MOF topology classification, void fraction prediction, and metal cluster/organic linker classification. This provides PMTransformer extensive knowledge of structure-property relationships in porous materials, which can be fine-tuned for specific adsorption properties. 

\begin{figure}[h!]
    \includegraphics[width=0.9\columnwidth]{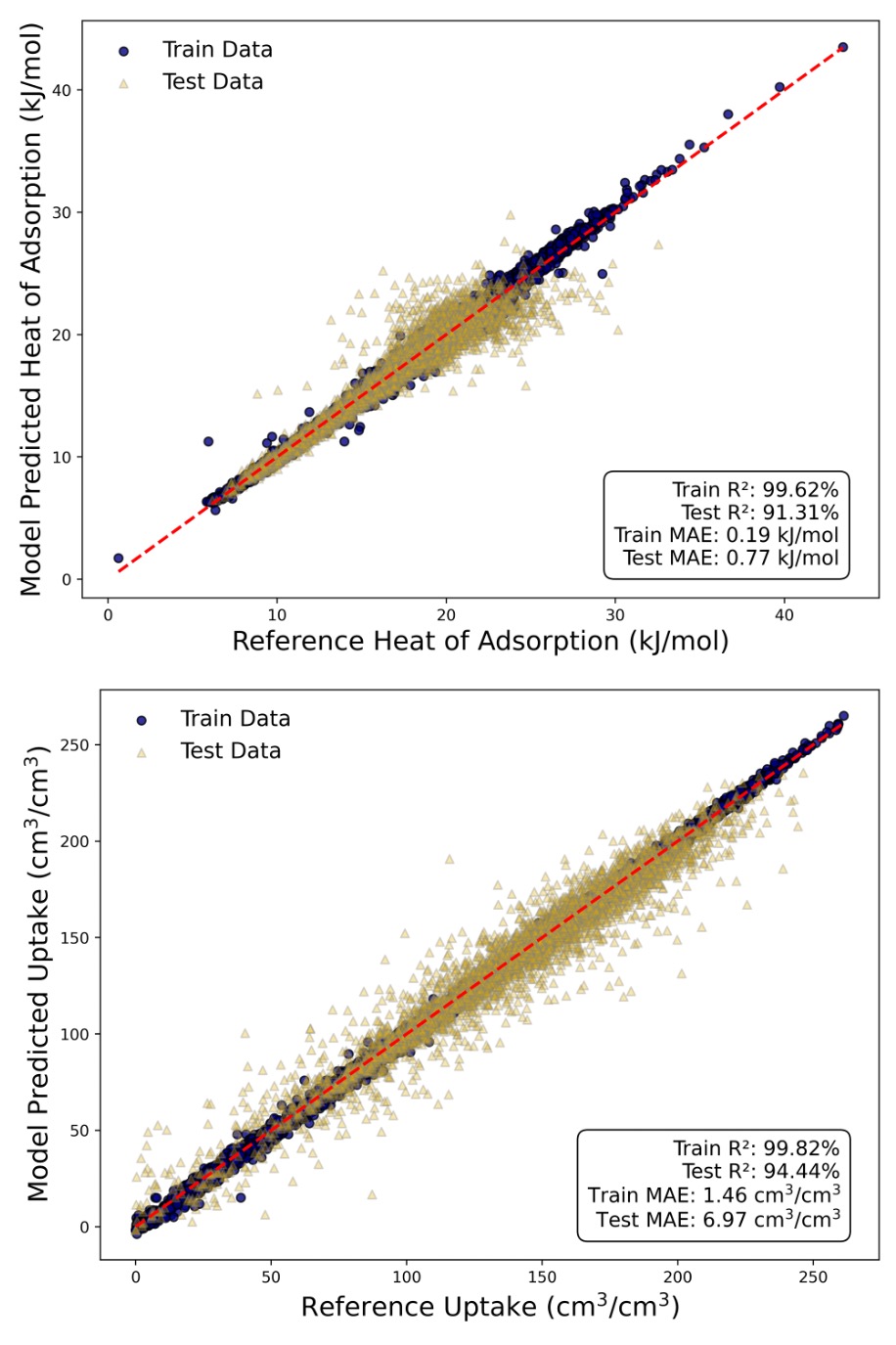} 
    \caption{Performance of the fine-tuned HoA model (upper panel, in units of kJ/mol) and the uptake model (lower panel, in units of cm$^3$/cm$^3$) against the reference data from the hMOF dataset. The training and testing data are shown in blue circles and yellow triangles, respectively. The dashed line represents the ideal 1:1 correlation. The MAE and R$^2$ values are shown in the lower right corner of each plot. }
    \label{fig:ml_parity}
\end{figure}

In this study, we fine-tuned two PMTransformer models, each to predict one of the two key methane adsorption properties: isosteric heat of adsorption and volumetric uptake at 298 K and 35 bar. We refer to these models as the HoA model and the uptake model, respectively, throughout the remainder of the paper.
The hypothetical MOF (hMOF) dataset~\cite{wilmer2012a} from the MOFX-DB database~\cite{bobbitt2023} was used for fine-tuning. 
The hMOF dataset contains hypothetical structures generated through systematic variation of MOF building blocks (metal nodes and organic linkers), with adsorption properties calculated using GCMC simulations. We selected this dataset because of its comprehensive coverage of structural diversity, consistent data sources, and availability of adsorption data for various gas types (e.g., CO$_2$, CH$_4$, H$_2$, Ar) and MOF metal centers (e.g., Zn, Cu, V, Zr).
From 151,464 structures that contain CH$_4$ adsorption data, we curated a subset of 32,768 structures through preprocessing. This curation involved removing structures with anomalous properties (e.g., zero surface area or missing adsorption values) and applying proportionate stratification based on surface area to ensure representative diversity across the structural space. The curated dataset was randomly divided into training, validation, and test sets with an 8:1:1 ratio. Our convergence analysis demonstrated that this subset size was sufficient to achieve satisfactory accuracy compared to the reference data, with mean absolute errors (MAE) of 0.77 kJ/mol for the HoA model and 6.97 cm$^3$/cm$^3$ (cm$^3$ of CH$_4$ at STP per cm$^3$ of MOF) for the volumetric uptake model (see Table S1 in the Supporting Information).
For brevity, the unit cm$^3$/cm$^3$ is used for all volumetric uptake values hereafter. 
Fig. \ref{fig:ml_parity} shows the training and testing performance of the two models, both trained on 80\% of 32,768 data points and tested on 10\%.
All adsorption data were standardized to have zero mean and unit variance before training. The fine-tuning process was performed with a batch size of 32 over 20 epochs, with the model's performance monitored through mean squared error (MSE) and mean absolute error (MAE) (see Figures S1 and S2 in the Supporting Information for the convergence of MSE/MAE values with respect to epochs).

After completing the fine-tuning process, we applied the optimized PMTransformer models to predict CH$_4$ heat of adsorption and volumetric uptake for the three metal-substituted MOFs studied in this work. Specifically, we generated predictions for M-HKUST-1, M-ATC, and M-ZIF-8 (M = Cu, Zn) using the DFT-relaxed structures as input to the models. This allows us to directly compare the effects of metal substitution on CH$_4$ adsorption predicted by machine learning and DFT methods.

\subsection{Monte Carlo Simulations}
\label{ssec:comput_details_mc}

Grand canonical Monte Carlo simulations were performed using the RASPA software package~\cite{dubbeldam2016} to provide reference adsorption data for evaluating our ML predictions and to investigate the underlying factors affecting prediction accuracy for metal-substituted MOFs. We conducted GCMC simulations for all six MOF structures in our study (M-HKUST-1, M-ATC, and M-ZIF-8, where M = Cu, Zn) as well as for a selected subset of hMOF structures with their metal-substituted variants. To maintain consistency with the GCMC parameters used in generating the hMOF dataset, we adopted simulation supercells consisting of 2$\times$2$\times$2 unit cells for all MOFs, with the MOF framework kept rigid. Methane molecules are modeled using the single-site TraPPE force field,~\cite{martin1998} while the parameters of MOF atoms are from the Universal Force Field (UFF).~\cite{rappe1992} Interactions between CH$_4$ and MOF atoms are described by Lennard-Jones (LJ) potentials after applying the Lorenz-Berthelot mixing rules.~\cite{lorentz1881ueber,berthelot1898} LJ cutoff distance of 12.8 \AA\, is applied.
To match the conditions used in the hMOF dataset, we performed GCMC simulations at 298 K and 35 bar. All simulations consists of 5000 initialization cycles followed by 5000 production cycles (higher than the 1500/1500 initialization/production cycles used in the hMOF dataset) to ensure convergence of the results. The GCMC simulations yield both CH$_4$ volumetric uptake and isosteric heat of adsorption values.

\section{Results and Discussions}
\label{sec:results}

\subsection{DFT results of CH$_4$ adsorption in M-MOFs}
\label{ssec:results_dft}

To establish a fundamental understanding of how metal substitution affects methane adsorption at the molecular level and set a baseline for the ML models, we first performed DFT calculations on the three original MOFs (Cu-HKUST-1, Cu-ATC, and Zn-ZIF-8) and their metal-substituted variants (Zn-HKUST-1, Zn-ATC, and Cu-ZIF-8). We focused on two strongest binding sites in each MOF, which have been identified in previous studies (for the original MOFs).~\cite{wuMetalOrganicFrameworks2010a,niu2019,fairen-jimenezFlexibilitySwingEffect2012} 

\begin{figure}[h!]
    \includegraphics[width=0.9\columnwidth]{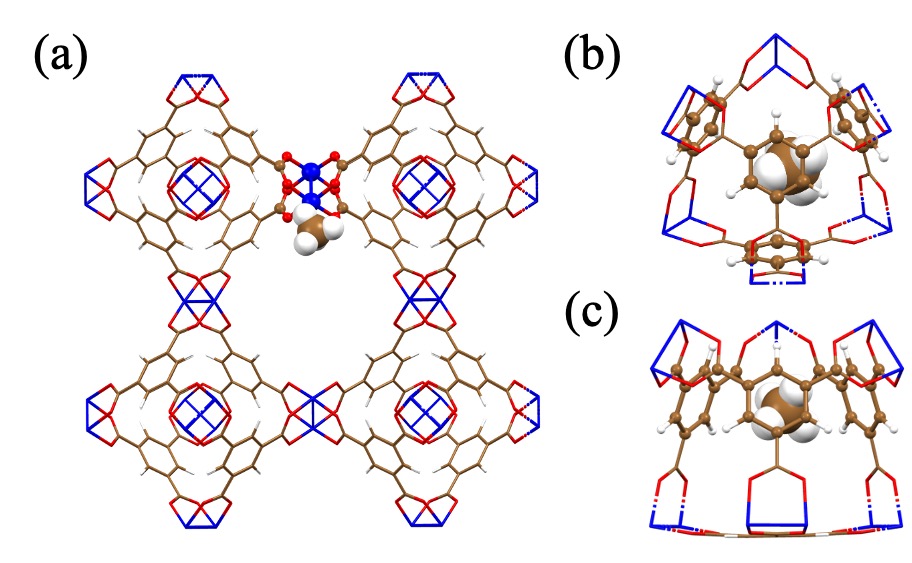} 
    \caption{Methane binding structures in Cu-HKUST-1 at site I (a) and site II (b), with the side view of site II in (c). See text for detailed description of the binding sites. For illustration purposes, the size of methane atoms is enlarged and atoms not associated with the binding sites are represented with wireframe models. Atom colors follow the scheme used in Fig. \ref{fig:bare_mofs}. The same applies to Figs. \ref{fig:atc} and \ref{fig:zif}.}
    \label{fig:hkust}
\end{figure}

The optimized binding geometries of methane in Cu-HKUST-1 are shown in Fig. \ref{fig:hkust}.
The first binding site (site I) is the coordinatively unsaturated Cu center in a paddlewheel building unit within the large 11 \AA\, cage (the yellow sphere in Fig. \ref{fig:bare_mofs}a).
At this site, methane preferentially adopts a configuration with three C-H bond oriented toward Cu to maximize the metal-methane interactions. The second binding site (site II, also referred to as a "window" site) is associated with the aromatic BTC linkers within the smaller 4 \AA\, cage (the purple sphere in Fig. \ref{fig:bare_mofs}a). Methane is trapped near a cage window due to enhanced van der Waals interactions with the aromatic rings. With neutron powder diffraction experiments, Wu et al.~\cite{wuMetalOrganicFrameworks2010a} reported that the open Cu site and the small cage window site in Cu-HKUST-1 are heavily populated, whereas other sites like the small cage center site and large cage corner site are only slightly populated.

\begin{figure}[h!]
    \includegraphics[width=0.9\columnwidth]{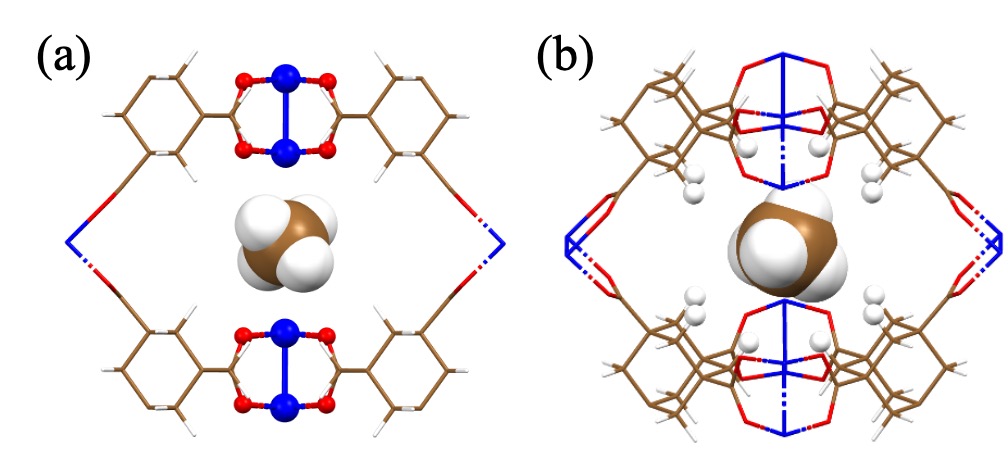} 
    \caption{Methane binding structures in Cu-ATC at site I (a) and site II (b).}
    \label{fig:atc}
\end{figure}

Similar to Cu-HKUST-1, Cu-ATC exhibits two primary binding sites, i.e. an open Cu site (site I) and a window site (site II), shown in Fig. \ref{fig:atc}.
Binding site I is located between the two opposing Cu paddlewheels that form nano-traps (the yellow sphere in Fig. \ref{fig:bare_mofs}b), allowing for enhanced Coulombic interaction between methane and both Cu centers.
Another cavity formed by the ATC linkers (the purple sphere in Fig. \ref{fig:bare_mofs}b) hosts site II, where methane interacts with twelve hydrogen atoms from the aliphatic hydrocarbons through van der Waals forces. The binding structures are consistent with those reported by Niu et al.~\cite{niu2019}

\begin{figure}[h!]
    \includegraphics[width=0.9\columnwidth]{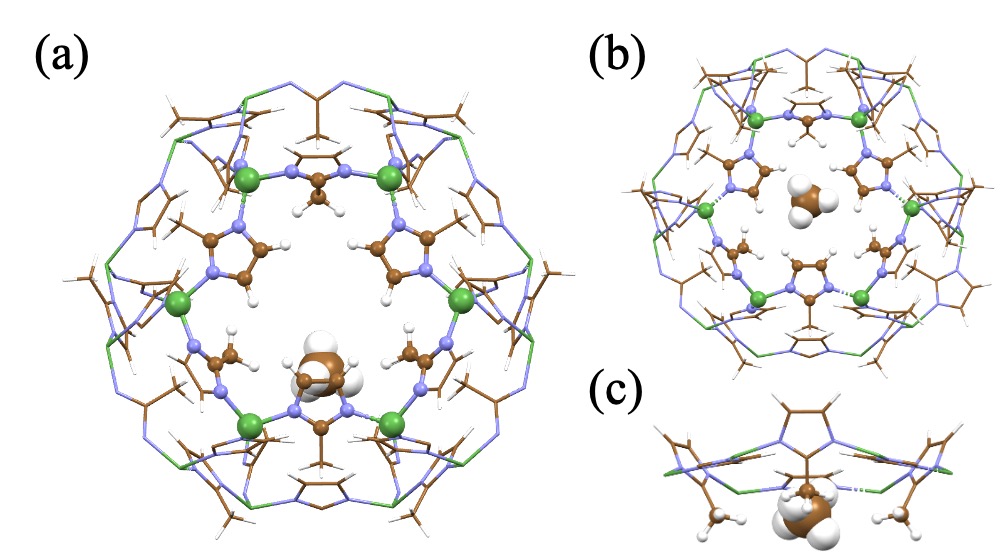} 
    \caption{Methane binding structures in Zn-ZIF-8 at site I (a) and site II (b), with the side view of site II in (c).}
    \label{fig:zif}
\end{figure}

Since all Zn centers in Zn-ZIF-8 are tetrahedrally coordinated to imidazolate linkers, there are no open metal sites available for methane binding. Instead, the two primary binding sites are both located near the 6-ring window (the blue dashed circle in Fig. \ref{fig:bare_mofs}c). The binding site I is associated with the C=C bond on the imidazolate ring, while site II is at the center of the 6-ring window due to interactions with the surrounding aromatic rings and methyl groups. The binding structures are shown in Fig. \ref{fig:zif}, consistent with those reported by Fairen-Jimenez et al.~\cite{fairen-jimenezFlexibilitySwingEffect2012}

Upon metal substitution (Cu-HKUST-1 $\rightarrow$ Zn-HKUST-1, Cu-ATC $\rightarrow$ Zn-ATC, and Zn-ZIF-8 $\rightarrow$ Cu-ZIF-8), the overall MOF topology and the locations of the two primary binding sites remain unchanged. 
M-HKUST-1 and M-ATC experience small expansion of the unit cell volume (3\%) while M-ZIF-8 experiences slighly more considerable contraction (-8.6\%). Such changes are expected due to the larger ionic radii of Zn$^{2+}$ compared to Cu$^{2+}$. Similar trends hold for the M-M (in the same paddlewheel unit) and M-O bond distances in M-HKUST-1 and M-ATC, as well as the tetrahedral M-N bond distances in M-ZIF-8, where the distances vary by 0.05-0.1 \AA\, between Cu and Zn variants.
Noticeable changes are found in the binding structure of site I for M-HKUST-1, where the M-CH$_4$ distance decreases by 0.45 \AA\, upon Cu$\rightarrow$Zn substitution. This indicates stronger binding of CH$_4$ to Zn compared to Cu, which is consistent with the higher adsorption energy calculated for Zn-HKUST-1 (see discussions next). Interestingly, the binding structure of site I in M-ATC shows a negligible change in the M-CH$_4$ distance (0.01 \AA) due to balanced attractions from the opposing metal centers. For non-metal binding sites (site II in M-HKUST-1 and M-ATC, sites I and II in M-ZIF-8), the binding structures are minimally changed, indicating that van der Waals interactions between CH$_4$ and organic linkers are not affected by metal substitution.
While this usually leads to similar adsorption energies for the two metal variants, it is not always the case (as we will see for M-ZIF-8). Geometric comparisons are summarized in Table S2 of the Supporting Information.

The calculated CH$_4$ adsorption energies at these binding sites are shown in Table \ref{tab:dft_energies}. For all MOFs except Cu-HKUST-1, our results show that site I exhibits stronger adsorption than site II.
For Cu-ATC and Zn-ZIF-8, our relative adsorption energies between the two sites agree well with previous DFT studies (see Table \ref{tab:dft_energies} for references). The numerical differences are likely due to the choice of DFT functional.
For Cu-HKUST-1, our results show that CH$_4$ binds stronger to site II than site I by 3.26 kJ/mol. This is consistent with previous experimental findings that CH$_4$ is preferentially adsorbed into the small cages (site II) first at low pressure~\cite{wu2015a} and the estimated CH4-small pore interactions was approximately -20 kJ/mol.~\cite{getzschmann2010}

\begin{table}[ht!]
    \caption{DFT calculated CH$_4$ adsorption energies (in kJ/mol) for the three MOFs and their metal-substituted variants at the two primary binding sites.}
    \label{tab:dft_energies}
    \begin{threeparttable}
    \begin{ruledtabular}
    \begin{tabular}{lcccc}
    & \multicolumn{2}{c}{This work} & \multicolumn{2}{c}{Literature values} \\
    \cmidrule{2-3} \cmidrule{4-5}
    MOF\tnote{a} & site I & site II & site I & site II \\
    \midrule
    Cu-HKUST-1\textsuperscript{*} & -17.15 & -20.41 & -14\tnote{b}  & -20\tnote{c} \\
    Zn-HKUST-1 & -34.62 & -21.16 & -24\tnote{b}  &  \\
    \addlinespace
    Cu-ATC\textsuperscript{*} & -31.80 & -23.28 & -28.99\tnote{d} & -23.67\tnote{d} \\
    Zn-ATC & -45.13 & -22.85 &  &  \\
    \addlinespace
    Zn-ZIF-8\textsuperscript{*} & -24.26 & -22.67 & -21.0\tnote{e} & -17.6\tnote{e} \\
    Cu-ZIF-8 & -17.91 & -13.13 &  &  \\
    \end{tabular}
    \end{ruledtabular}
    \begin{tablenotes}[flushleft]
    \item[a] The original MOFs, denoted by asterisks, are Cu-HKUST-1, Cu-ATC, and Zn-ZIF-8. The metal-substituted variants are Zn-HKUST-1, Zn-ATC, and Cu-ZIF-8.
    \item[b] Ref. \citenum{kohPredictingMethaneStorage2015} (vdW-DF2). 
    \item[c] Ref. \citenum{getzschmann2010} (experiment).
    \item[d] Ref. \citenum{niu2019} (vdW-DF2).
    \item[e] Ref. \citenum{fairen-jimenezFlexibilitySwingEffect2012} (PBE-D2).
    \end{tablenotes}
    \end{threeparttable}
\end{table}

Metal substitution significantly affects the CH$_4$ adsorption energy in all three MOFs. 
For binding sites featuring open metal centers (site I in M-HKUST-1 and M-ATC), we would expect strong dependence of the binding energy on the choice of metal.
Indeed, for M-HKUST-1, we see a large difference between site I of Cu-HKUST-1 (-17.15 kJ/mol) and that of Zn-HKUST-1 (-34.62 kJ/mol), with the Zn variant exhibiting much stronger CH$_4$ binding. Similarly, for M-ATC, the Zn variant shows stronger binding in site I (-45.13 kJ/mol) compared to the original Cu-ATC (-31.80 kJ/mol).
For site II of these two MOFs, due to the lack of open metal centers, the binding energies of the Cu and Zn variants are similar (differ by $\leq$ 1 kJ/mol), yielding an overall stronger binding energy for the Zn-substituted MOFs.
This trend is consistent with previous studies on open-metal-based M-MOFs, such as M-HKUST-1 and M-MOF-74, that have shown higher methane uptake in the Zn variant compared to its Cu analog.~\cite{kohPredictingMethaneStorage2015,lee2015b}
Surprisingly, the Zn variant of M-ZIF-8 shows a significantly stronger adsorption energy in both sites I and II (-24.26 kJ/mol and -22.67 kJ/mol, respectively) compared to the Cu variant (-17.91 kJ/mol and -13.13 kJ/mol, respectively). This is somewhat unexpected, as M-ZIF-8 does not feature open metal sites, and both binding sites are located near the organic linkers. Since no structural distorsions are observed near the binding sites, we attribute this difference to the indirect electronic effects caused by metal substitution. Further investigation is warranted to understand the underlying mechanism of this phenomenon.

To summarize this section, our DFT results establish that metal substitution significantly affects methane binding strength in all three MOFs, with or without open metal sites. Specifically, the Zn variants always exhibit stronger binding than the Cu variants. These DFT predictions set the stage for evaluating how effectively machine learning models can capture metal-substitution effects when predicting macroscopic adsorption properties.

\subsection{Machine learning predictions and error analysis}
\label{ssec:results_ml}

\begin{figure}[h!]
    \includegraphics[width=0.9\columnwidth]{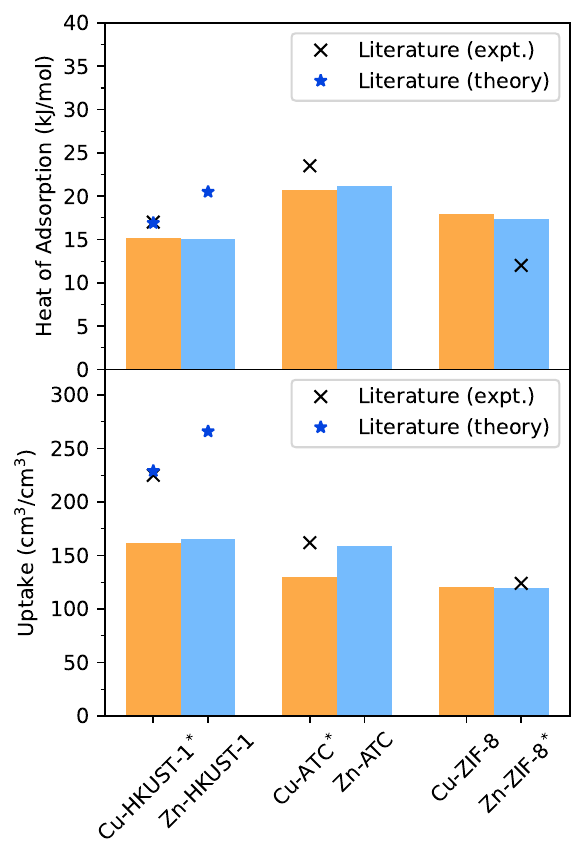} 
    \caption{Performance of fine-tuned PMTransformer models on predicting CH$_4$ heat of adsorption (upper panel, in units of kJ/mol) and volumetric uptake (lower panel, in units of cm$^3$/cm$^3$) for the three MOFs and their metal-substituted variants. Experimental literature data from refs \citenum{mason2014,niu2019,zhou2007} are shown as crosses and theoretical predictions from ref \citenum{kohPredictingMethaneStorage2015} (GCMC simulations based on DFT-derived force fields) are shown as stars. Cu-ATC's experimental uptake at 298 K and 35 bar is estimated using parameters from Niu et al.~\cite{niu2019}, who reported a dual-site Langmuir-Freundlich model fitted to their measured isotherms (298 K, up to 1 bar). The original MOFs are marked by asterisks.}
    \label{fig:ml_vs_expt}
\end{figure}

Having established the impact of metal substitution on methane binding energetics through DFT calculations, we now evaluate the performance of our two fine-tuned ML models in predicting macroscopic CH$_4$ adsorption properties. Fig. \ref{fig:ml_vs_expt} shows the predicted heat of adsorption and volumetric uptake values for M-HKUST-1, M-ATC, and M-ZIF-8 (M = Cu, Zn) using the corresponding fine-tuned PMTransformer models.
The predicted heat of adsorption values (by the HoA model) are 15.2, 20.7, and 17.9 kJ/mol for Cu-HKUST-1, Cu-ATC, and Zn-ZIF-8, respectively, with an MAE of 3.5 kJ/mol compared to the experimental values.
Note that while DFT adsorption energies are reported as negative values (where more negative indicates stronger binding), heat of adsorption values are conventionally reported as positive (where more positive indicates stronger binding).
The volumetric uptake values (predicted by the uptake model) are 161.5, 129.6, and 120.8 cm$^3$/cm$^3$, respectively, with an MAE of 33 cm$^3$/cm$^3$ compared to experiment.
The ML errors are higher than those observed in the testing sets (0.77 kJ/mol and 6.97 cm$^3$/cm$^3$), which is expected since the test set is a subset of the hMOF dataset, while the experimental data are from different sources. 
Despite this, the ML models correctly capture the experimental trends that (1) Cu-ATC yields the highest heat of adsorption and (2) Cu-HKUST-1 $>$ Cu-ATC $>$ Zn-ZIF-8 for uptake capacity at 298 K and 35 bar.

However, a striking pattern emerges when we examine the predicted values for the metal-substituted variants. The ML models predict nearly identical adsorption properties for Cu and Zn variants of each MOF, which is in contrast to the DFT results that show significant differences in methane binding strength between the two metal types. For M-HKUST-1, despite the 11 kJ/mol difference in DFT calculated adsorption energies (site I), the heat of adsorption values predicted by the HoA model differ by only 0.2 kJ/mol between M = Cu and M = Zn. Similarly, for M-ATC and M-ZIF-8, the model predicts less than 1 kJ/mol difference in heat of adsorption between the two variants. 
For volumetric uptake, Koh et al. reported a notably higher value for Zn-HKUST-1 (266 cm$^3$/cm$^3$) compared to Cu-HKUST-1 (229 cm$^3$/cm$^3$) at 298 K and 35 bar using GCMC simulations with vdW-DF2 derived force fields,~\cite{kohPredictingMethaneStorage2015} consistent with our DFT results. In contrast, the uptake model predicts a negligible difference of 4.1 cm$^3$/cm$^3$ between the two variants. The predicted uptake values for M-ATC and M-ZIF-8 are also similar, with differences of 11.1 cm$^3$/cm$^3$ and 1.2 cm$^3$/cm$^3$, respectively.

To investigate the source of this discrepancy, we conducted GCMC simulations using the UFF/TraPPE force field combination{\textemdash}the same computational method used to generate the adsorption data for the hMOF database on which our ML models were fine-tuned. These GCMC simulations predict minimal differences between Cu and Zn variants across all three MOFs (see Table \ref{tab:ml_vs_gcmc}), mirroring the pattern observed in our ML predictions. For heat of adsorption, the GCMC results differ by 0.4, 0.1, and 1.4 kJ/mol for M-HKUST-1, M-ATC, and M-ZIF-8, respectively. For volumetric uptake, the differences are 12.2, 3.4, and 8.1 cm$^3$/cm$^3$, respectively. The ML predictions align with their GCMC references very closely, with average absolute errors of 0.6 kJ/mol for heat of adsorption and 15.7 cm$^3$/cm$^3$  for volumetric uptake. Such consistency between ML predictions and UFF/TraPPE-based GCMC results strongly suggests that the models' inability to distinguish metal variants stems not from algorithmic limitations but from the training data itself. 

\begin{table}[ht]
    \caption{Comparison of fine-tuned PMTransformer models and GCMC simulations for heat of adsorption (kJ/mol) and volumetric uptake (cm$^3$/cm$^3$) for the three MOFs under study, the four randomly selected hypotential MOFs, and their metal-substituted variants.}
    \label{tab:ml_vs_gcmc}
    \begin{threeparttable}
    \begin{ruledtabular}
    \begin{tabular}{lcccc}
    & \multicolumn{2}{c}{Heat of Adsorption} & \multicolumn{2}{c}{Uptake} \\
    \cmidrule{2-3} \cmidrule{4-5}
    MOF\tnote{a} & ML\tnote{b} & GCMC\tnote{d} & ML\tnote{c} & GCMC\tnote{d} \\
    \midrule
    Cu-HKUST-1\textsuperscript{*} & 15.2 & 15.3 & 161.5 & 178.1 \\
    Zn-HKUST-1 & 15.0 & 15.7 & 165.6 & 190.3 \\
    \addlinespace
    Cu-ATC\textsuperscript{*} & 20.7 & 21.5 & 129.6 & 155.4 \\
    Zn-ATC & 21.2 & 21.6 & 140.7 & 158.8 \\
    \addlinespace
    Cu-ZIF-8 & 17.4 & 17.7 & 119.6 & 125.7 \\
    Zn-ZIF-8\textsuperscript{*} & 17.9 & 16.3 & 120.8 & 117.6 \\
    \addlinespace
    Cu-hMOF-32693\textsuperscript{*} & 21.5 & 21.6 & 180.5 & 153.7\\
    Zn-hMOF-32693 & 21.7 & 22.0 & 181.8 & 151.5\\
    \addlinespace
    Cu-hMOF-5041020 & 18.6 & 16.6 & 160.1 & 134.1\\
    Zn-hMOF-5041020\textsuperscript{*} & 18.9 & 16.8 & 165.0 & 136.6\\
    \addlinespace
    Cu-hMOF-5052408 & 14.9 & 14.6 & 133.5 & 129.3\\
    Zn-hMOF-5052408\textsuperscript{*} & 14.9 & 15.1 & 136.1 & 130.1\\
    \addlinespace
    Cu-hMOF-5073729\textsuperscript{*} & 13.9 & 13.9 & 177.5 & 190.3\\
    Zn-hMOF-5073729 & 14.0 & 14.0 & 177.1 & 191.7\\
    \end{tabular}
    \end{ruledtabular}
    \begin{tablenotes}
        \item[a] The original unaltered MOFs are marked by asterisks.
        \item[b] The HoA model is used for heat of adsorption predictions.
        \item[c] The uptake model is used for volumetric uptake predictions. 
        \item[d] GCMC simulations based on the UFF/TraPPE force fields are performed with similar parameters as those used in the hMOF database (more details in Section \ref{ssec:comput_details_mc}).
    \end{tablenotes}
    \end{threeparttable}
\end{table}
        
This outcome is not surprising given that UFF treats different metals through basic parameters like atomic radii and electronegativity without capturing the complex electronic effects, particularly at open metal sites. It has been well documented that generic force fields such as UFF are not suitable for accurately describing the interactions at open metal sites in MOFs.~\cite{grajciar2010,dzubak2012,kohPredictingMethaneStorage2015,yan2022a}
For instance, in ref. \citenum{kohPredictingMethaneStorage2015}, the authors noted that the UFF/TraPPE force fields are not able to predict the differences in binding energy between variants of metal-substituted HKUST-1, including Cu-HKUST-1 and Zn-HKUST-1. To further validate our finding, we randomly selected 4 structures from the hMOF database (see Figure S3 in the Supporting Information for their structures) and created metal-substituted variants by replacing Cu (Zn) with Zn (Cu). GCMC simulations using UFF/TraPPE for these 8 structures (4 original MOFs and 4 metal-substituted variants) showed minimal differences in adsorption properties, with an average absolute difference of just 0.7 kJ/mol in heat of adsorption and 5.8 cm$^3$/cm$^3$ in uptake between metal variants (see Table \ref{tab:ml_vs_gcmc}). When trained on this data, even a perfect ML model would inevitably fail to distinguish the effects of metal substitution.

\section{Conclusions}
\label{sec:conclusions}
In this study, we performed computational investigation of methane adsorption in metal-substituted MOFs using density functional theory and machine learning approaches. Our DFT calculations revealed significant differences in methane binding energetics between Cu and Zn variants of all three MOFs under study. For M-HKUST-1 and M-ATC, metal substitution from Cu to Zn results in a 11-13 kJ/mol increase in binding energy at site I (the open metal site), while the binding energy at site II (the window site) remains nearly unchanged (difference $\leq$ 0.5 kJ/mol). Surprisingly, for M-ZIF-8 which does not contain open metal sites, the binding energy at both sites I and II is significantly stronger for the Zn variant compared to the Cu variant, with a difference of 6-7 kJ/mol.

While our fine-tuned PMTransformer models predict qualitatively correct methane adsorption properties for the original MOF structures, they fail to distinguish between metal variants despite the significant differences identified by DFT. Through further investigation, we determined that this limitation stems from the reference data used for training. GCMC simulations using the UFF/TraPPE force field combination{\textemdash}the same method employed to generate the hMOF training database{\textemdash}predict nearly identical adsorption properties for Cu and Zn variants, mirroring the pattern observed in our ML predictions. 

These findings highlight a critical limitation in current ML-based computational screening of MOFs: while ML models like PMTransformer can accurately predict adsorption properties within the constraints of their training data, they inherit any systematic biases present in the data. For metal-substituted MOFs, generic force fields such as UFF used in high-throughput GCMC simulations fail to capture the electronic effects that significantly influence methane binding at open metal sites, leading to ML models that cannot effectively distinguish between different metal types, despite their demonstrated capability to predict other structural effects on adsorption.
For applications where electronic effects significantly impact adsorption properties, this presents a fundamental challenge that must be addressed through improved reference data. 
To date, specialized force fields based on DFT/wavefunction theory~\cite{kohPredictingMethaneStorage2015,haldoupis2015,becker2018,li2024} and machine learning potentials derived from DFT calculations~\cite{vandenhaute2023,liu2024a,wieser2024,sharma2024} have been developed to address this issue. 
However, these approaches are not widely available in high-throughput simulations due to the costly DFT calculations required to generate the reference data, which limits their applicability in large-scale screening of MOFs.
Future work may include creating hybrid datasets that combine high-throughput GCMC results with DFT calculations at metal binding sites. These advancements would accelerate the discovery of optimal MOFs for methane capture and storage through more accurate computational screening.

\begin{acknowledgments}
We thank Yat Li for helpful discussions. X.W. acknowledges the start-up funding from the University of California, Santa Cruz (UCSC). Computational resources were provided by the UCSC Hummingbird supercomputer. 
\end{acknowledgments}

\bibliography{references}% Add references in references.bib

\end{document}